\documentclass[a4paper,11pt]{article}

\usepackage{jheppub} 
\usepackage[normalem]{ulem}  

\newcommand{\Ga}{{\Gamma}}
\newcommand{\De}{{\Delta}}

\newcommand{\Om}{{\Omega}}

\newcommand{\al}{{\alpha}}
\newcommand{\bt}{{\beta}}
\newcommand{\ga}{{\gamma}}
\newcommand{\de}{{\delta}}
\newcommand{\ep}{{\epsilon}}

\newcommand{\te}{{\theta}}

\newcommand{\lm}{{\lambda}}

\newcommand{\sig}{{\sigma}}

\newcommand{\bsa}{{\boldsymbol{a}}}

\newcommand{\bsk}{{\boldsymbol{k}}}

\newcommand{\bsw}{{\boldsymbol{w}}}
\newcommand{\bsx}{{\boldsymbol{x}}}

\newcommand{\bsz}{{\boldsymbol{z}}}

\newcommand{\bsep}{{\boldsymbol{\epsilon}}}

\newcommand{\bsnl}{{\boldsymbol{0}}}

\newcommand{\bse}{{\boldsymbol{e}}}

\newcommand{\bska}{{\boldsymbol{\kappa}}}

\newcommand{\bbY}{{\mathbb{Y}}}

\newcommand{\cA}{{\mathcal{A}}}

\newcommand{\cG}{{\mathcal{G}}}

\newcommand{\lan}{{\langle}}

\newcommand{\nn}{{\nonumber}}
\newcommand{\ol}{\overline}
\newcommand{\pd}{{\partial}}

\newcommand{\ran}{{\rangle}}

\newcommand{\wh}{\widehat}

\def\bbra{{\langle\kern-2.5pt\langle}}
\def\kket{{\rangle\kern-2.5pt\rangle}}

\title{Momentum space approach to crossing symmetric CFT correlators II: General spacetime dimension}

\author[a]{Hiroshi Isono,}
\author[b]{Toshifumi Noumi,}
\author[c]{and Gary Shiu}

\affiliation[a]{Department of Physics, Faculty of Science, Chulalongkorn University, Bangkok 10330, Thailand}
\affiliation[b]{Department of Physics, Kobe University, Kobe 657-8501, Japan}
\affiliation[c]{Department of Physics, University of Wisconsin-Madison, Madison, WI 53706, USA}

\emailAdd{hiroshi.isono81@gmail.com}
\emailAdd{tnoumi@phys.sci.kobe-u.ac.jp}
\emailAdd{shiu@physics.wisc.edu}

\preprint{KOBE-COSMO-19-13, MAD-TH-19-06}

\abstract{
Our previous work
\cite{Isono:2018rrb} constructed, in three-dimensional momentum space, a manifestly crossing symmetric basis for scalar conformal four-point functions, based on the factorization property proposed by Polyakov. This work extends this construction to general dimensional conformal field theory. To facilitate the treatment of symmetric traceless tensors, we exploit techniques of spherical harmonics in general dimensions. 
}

\begin{document} 
\setcounter{tocdepth}{2}
\maketitle
\flushbottom

\section{Introduction}
\setcounter{equation}{0}

The crossing symmetric basis of conformal four-point functions, pioneered by Polyakov in 1974 \cite{Polyakov:1974gs}, is based on the following ansatz of the expansion of the four-point functions\footnote{Primed correlators are defined as $\lan\,...\,\ran=(2\pi)^d\delta^{(d)}(\sum_i\bsk_i)\lan\,...\,\ran'$. },
\begin{align}
\langle O_1(\bsk_1)O_2(\bsk_2)O_3(\bsk_3)O_4(\bsk_4)\rangle'
=\sum_O\left(W_O^{(\textbf s)}+W_O^{(\textbf t)}+W_O^{(\textbf u)}\right)+\text{(analytic terms)}\,,
\end{align}
where the index $O$ labels the intermediate primary operators. 
$W_O^{(\textbf s)}$ is called the Polyakov block in the $\textbf s$-channel. Each block satisfies the consistent factorization in momentum space, just as the on-shell factorization of scattering amplitudes of ordinary field theories. The Polyakov block with external scalar operators was shown in Mellin space to be nothing but the Witten exchange diagram \cite{Gopakumar:2016wkt,Gopakumar:2016cpb}. 
Our paper \cite{Isono:2018rrb} showed directly in three-dimensional momentum space that the Witten exchange diagram is a natural consequence of the consistent factorization,
and constructed the Polyakov block with an intermediate symmetric traceless operator of general spin.

\medskip
This paper extends the construction in \cite{Isono:2018rrb} to general dimensions. 
From a technical point of view, the extension to general dimensions becomes a bit more involved due to the treatment of symmetric traceless tensors of general spins. In the three dimensional case \cite{Isono:2018rrb}, we adopted the helicity representation instead of dealing with explicit vector indices. More concretely, one first fixes one of the momenta in the correlation function by using rotational symmetry without loss of generality. One then decomposes a symmetric traceless tensor in the correlator into irreducible representations of the little group of the fixed momentum. In three-dimension, the little group is ${\rm O}(2)$ and the expansion is nothing but the Fourier expansion \cite{Arkani-Hamed:2015bza,Isono:2018rrb}. In general dimension, the Fourier expansion is replaced by the expansion in spherical harmonics on the general dimensional unit sphere. In the present paper we elucidate this point in general spacetime dimension to construct a crossing-symmetric basis of scalar four-point functions.

\medskip

The outline of the rest of this paper is as follows. Sec.~\ref{sec:harmonic} is a brief review of the harmonic analysis needed for our analysis. In Sec.~\ref{sec:correlator} we find the helicity representation of two- and three-point functions with spins, and discuss their analytic properties to define the cubic vertex. In Sec.~\ref{sec:crossing} we construct the crossing symmetric basis of scalar four-point functions. The two appendices present derivations of some integral formulae in Sec.~\ref{sec:correlator}.

\section{Properties of spherical harmonics}
\label{sec:harmonic}

This section reviews basic properties of spherical harmonics in general spacetime dimensions.
The machinery will play a main role in expanding scalar functions of momenta and polarization vectors which appear in conformal two- and three-point functions. Readers familiar with spherical harmonics in general dimension may jump safely to Sec.~\ref{sec:correlator} after checking the Funk-Hecke formula introduced in Sec.~\ref{subsec:FH}

\paragraph{${\rm SO}(2)$ spherical harmonics.}
Before going into the general dimensional case, let us consider lower dimensional cases, in which the expansion is achieved easily.
As the simplest case, we begin with a scalar function of two unit vectors, $\hat\bsw=(\hat w_1,\hat w_2)$ and $\hat\bsz=(\hat z_1,\hat z_2)$, in two dimension.  More explicitly, it is a function of the inner product of the two unit vectors, $f(\hat\bsw\cdot\hat\bsz)$. If this function is regular at $\hat\bsw\cdot\hat\bsz=0$, we may expand it as 
\begin{align}
f(\hat\bsw\cdot\hat\bsz)
&=\lambda_0+\sum_{m=1}^\infty 
\lambda_{m}\Big[(\hat w_1+i\hat w_2)^m(\hat z_1-i\hat z_2)^m
+(\hat w_1-i\hat w_2)^m(\hat z_1+i\hat z_2)^m\Big]\,,
\label{SO(2)_expansion}
\end{align}
where $\lambda_m$ are constant parameters. Notice that the functions $(\hat z_1\pm i\hat z_2)^m$ are nothing but the ${\rm SO}(2)$ spherical harmonics. If we denote a basis of the ${\rm SO}(2)$ spherical harmonics by
\begin{align}
Y_0(\hat\bsz)=1\,,
\quad
Y_{m+}(\hat\bsz)=(\hat z_1+i\hat z_2)^m\,,
\quad
Y_{m-}(\hat\bsz)=(\hat z_1-i\hat z_2)^m\,,
\end{align}
we may rewrite Eq.~\eqref{SO(2)_expansion} as
\begin{align}
\label{SO(2)_Pi}
f(\hat\bsw\cdot\hat\bsz)=\sum_{m=0}^\infty 
\lambda_{m}\Pi_m(\hat\bsw,\hat\bsz)
\,,
\end{align}
where we introduced the projector $\Pi_m$ onto the spin $m$ sector as
\begin{align}
\Pi_0(\hat\bsw,\hat\bsz)&=Y_0(\hat\bsw)Y_0^*(\hat\bsz)\,,
\\
\Pi_m(\hat\bsw,\hat\bsz)&=
Y_{m+}(\hat\bsw)Y_{m+}^*(\hat\bsz)
+
Y_{m-}(\hat\bsw)Y_{m-}^*(\hat\bsz)
\quad(m=1,2,\ldots)
\,.
\end{align}
The expansion is nothing but the Fourier expansion, as is manifest in polar coordinates.

\paragraph{${\rm SO}(3)$ spherical harmonics.}

A similar expansion applies in the three dimensional case. As we physicists are familiar with, any regular scalar function of two unit vectors, $\hat\bsw$ and $\hat\bsz$, in three dimension may be expanded by the Legendre polynomials $P_m(\hat\bsw\cdot\hat\bsz)$. More explicitly, we may write it in the form~\eqref{SO(2)_Pi} with some $\Pi_m\propto P_m(\hat\bsw\cdot\hat\bsz)$. Since the addition theorem of the ${\rm SO}(3)$ spherical harmonics states that for any orthonormal basis $Y_{mn}$,
\begin{align}
P_m(\hat\bsw\cdot\hat\bsz)\propto \sum_{n=-m}^mY_{mn}(\hat\bsw)Y_{mn}^*(\hat\bsz)\,,
\end{align}
we may normalize $\Pi_m$ such that
\begin{align}
\Pi_m(\hat\bsw,\hat\bsz)=\sum_{n=-m}^mY_{mn}(\hat\bsw)Y_{mn}^*(\hat\bsz)\,,
\end{align}
which is again nothing but the projector onto the spin $m$ sector.

\medskip

\subsection{Spherical harmonics in general dimension and Funk-Hecke formula}
\label{subsec:FH}
We can generalize the expansion of scalar functions with two unit vectors to general dimensions in a straightforward manner. The expansion is called the Funk-Hecke formula~\cite{Erdelyi2,Vilenkin,AH}: Any scalar function of two unit vectors, $\hat\bsw$ and $\hat\bsz$, in $D$ dimension is expanded as
\begin{align}
f(\hat\bsw\cdot\hat\bsz)=\sum_{m=0}^\infty 
\lambda_{m}\Pi_m(\hat\bsw,\hat\bsz)
\quad
{\rm with}
\quad
\Pi_m(\hat\bsw,\hat\bsz)=\sum_{n=1}^{\dim\bbY^D_m}Y_{mn}(\hat\bsw)Y_{mn}^*(\hat\bsz)
\,,
\label{FH1}
\end{align}
where $Y_{mn}$ stands for an orthonormal basis of ${\rm SO}(D)$ spherical harmonics with total spin $m$ and $\dim\bbY^D_m$ is the dimension of the spin $m$ representation of ${\rm SO}(D)$\footnote{
In more detail, spherical harmonics are elements of the eigenspace $\bbY^D_m$ with eigenvalues $m(m+D-2)$ of the quadratic Casimir operator of ${\rm SO}(D)$,
\begin{align}
\Om_2 = -\frac{1}{2}(z^\mu\pd^\nu - z^\nu\pd^\mu)(z_\mu\pd_\nu-z_\nu\pd_\mu) \, ,
\end{align}
where $z^\mu$ is a general $D$-component vector satisfying $z^\mu = |\bsz| \hat{z}^\mu$. Note that $\Om_2$ is independent of $|\bsz|$.
The non-negative integer $m$ is the total spin. The dimension of $\bbY^D_m$ is given by
\begin{align}
\dim \bbY^D_m &= \frac{(m+D-1)!}{m!(D-1)!} - \frac{(m+D-3)!}{(m-2)!(D-1)!} \,.
\label{dimYds}
\end{align} 
See also the footnote in Appendix~\ref{V3m-proof} for the relation to harmonic polynomials.
}.
The coefficients $\lambda_m$ are evaluated by the integral,
\begin{align}
\lm_m
&= 
\frac{{\rm vol}(S^{D-2})}{{\rm vol}(S^{D-1})} \int_{-1}^1 dt \, (1-t^2)^{\frac{D-3}{2}} P^{(D)}_{m}(t) f( t )\,,
\label{FH2}
\end{align}
where $P^{(D)}_{m}(t)$ is the normalized Gegenbauer polynomial defined by
\begin{align}
P^{(D)}_{m}(t) := \frac{C^{({D}/{2}-1)}_m(t)}{C^{({D}/{2}-1)}_m(1)}
\label{normalized_G}
\end{align}
with the Gegenbauer polynomial,
\begin{align}
C_m^{(\al)}(t) = \sum_{n=0}^{[m/2]} (-)^n \frac{\Ga(m-n+\al)}{n!(m-2n)!\Ga(\al)} (2t)^{m-2n}\,, \qquad
C_m^{(\al)}(1) = \frac{(2\al)_m}{m!}\,. 
\end{align}
Here $[x]$ is the integer part of $x$ and $(\alpha)_m=\Gamma(\alpha+m)/\Gamma(\alpha)$ is the shifted factorial (also known as the Pochhammer symbol). We may also write the projector $\Pi_m$ as\footnote{We will not attach the dimensionality $D$ to the projector $\Pi_m$ and the orthonormal basis vectors $Y_{mn}$ as long as it is obvious from the context.
}
\begin{align}
\Pi_m(\hat\bsw,\hat\bsz)= \dim\bbY^D_m \cdot P^{(D)}_{m}(\hat\bsw \cdot \hat\bsz)\,.
\label{projector}
\end{align}
This is called the addition theorem of spherical harmonics.
In the rest of this section we summarize the basic properties of the spherical harmonics and review the derivation of the Funk-Hecke formula.

\subsection{Derivation of the Funk-Hecke formula}

We begin with the fact that any function on the unit sphere $S_{D-1}$ can be expanded in spherical harmonics,
\begin{align}
f(\hat\bsz) = \sum_{m=0}^\infty \sum_{n=1}^{\dim\bbY^D_m} c_{mn}Y_{mn}(\hat\bsz) \,.
\label{funcdecomp}
\end{align}
The basis spherical harmonics $Y_{mn}$ are orthogonal and normalized as
\begin{align}
\lan Y_{mn}, Y_{m'n'} \ran = \de_{mm'}\de_{nn'} 
\end{align}
with the inner product defined by
\begin{align}
\lan f, g \ran = \int_{\hat\bsz} d\sig_{D-1} \, f^*(\hat\bsz) g(\hat\bsz)\,.
\end{align}
Here the integration measure $d\sig_{D-1}$ is the standard one on the unit sphere $S_{D-1}$ with the normalization,
\begin{align}
\int_{\hat\bsz} d\sig_{D-1} \, 1 = 1\,.
\end{align}
We then introduce the projection operator onto the spin $m$ subspace $\bbY^D_m$ as
\begin{align}
\label{PiYY}
\Pi_{m}(\hat\bsz, \hat\bsw) &= \sum_{n=1}^{\dim\bbY^D_m} Y_{mn}(\hat\bsz) Y^*_{mn}(\hat\bsw) \,.
\end{align}
In terms of the projection operator, the spin decomposition \eqref{funcdecomp} reads
\begin{align}
\label{Pif}
f = \sum_{m=0}^\infty f_m \, , \qquad
f_m(\hat\bsz) = \int_{\hat\bsw} d\sig_{D-1} \, \Pi_m(\hat\bsz, \hat\bsw) f(\hat\bsw) \, .
\end{align}

\medskip

\paragraph{Explicit form of the projector.}
Let us derive the expression~\eqref{projector} of the projector $\Pi_m$ 
{exploiting its rotational invariance.}
For this purpose, we first introduce the little group ${\rm O}(D,\hat\bsa)$ of a unit vector $\hat\bsa$ by
\begin{align}
{\rm O}(D,\hat\bsa) := \{A \in {\rm O}(D): ~ A\hat\bsa=\hat\bsa\} \,.
\end{align}
A function $f(\hat\bsz)$ on $S_{D-1}$ is called ${\rm O}(D,\hat\bsa)$-invariant when $f(A\hat\bsz)=f(\hat\bsz)$ holds for any element $A \in {\rm O}(D,\hat\bsa)$.
It is known that any ${\rm O}(D,\hat\bsa)$-invariant spherical harmonic of spin $m$ is proportional to the normalized Gegenbauer polynomial~\cite{AH},
\medskip
\begin{align}
\label{ODainv}
 f (\hat\bsz)\in \bbY^{D}_m \mbox{ is } {\rm O}(D,\hat\bsa) \mbox{-invariant } 
\Longleftrightarrow \,
f(\hat\bsz) = P^{(D)}_{m}(\hat\bsz \cdot \hat\bsa)f(\hat\bsa) ~
\mbox{ for any } \hat\bsz \in S^{D-1} \, .
\end{align}
We will often use this property. As the first example,
let us prove the addition theorem \eqref{projector}.
Notice first that the projector $\Pi_m(\hat\bsz, \hat\bsw)$ is ${\rm O}(D)$-invariant because 
the rotated spherical harmonics $Y_{mn}(A\hat\bsz)$ $(A\in {\rm O}(D))$ form another orthonormal basis of spin $m$. 
In particular, we have $\Pi_m(A\hat\bsz, \hat\bsw)=\Pi_m(\hat\bsz, \hat\bsw)$ for any little group transformation $A\in {\rm O}(D,\hat\bsw)$. Therefore, if we think of the projector $\Pi_m(\hat\bsz, \hat\bsw)$ as a function of $\hat\bsz$, it is an ${\rm O}(D,\hat\bsw)$-invariant spherical harmonic of spin $m$. Therefore, it follows from the equivalence \eqref{ODainv} that for each $m$
\begin{align}
\Pi_m(\hat\bsz, \hat\bsw)=\Pi_m(\hat\bsw, \hat\bsw)P^{(D)}_{m}(\hat\bsz \cdot \hat\bsw)=\dim\bbY^D_m \cdot P^{(D)}_{m}(\hat\bsz \cdot \hat\bsw)\,.
\end{align}

\medskip

\paragraph{The Funk-Hecke formula.}
The Funk-Hecke formula \eqref{FH1} with \eqref{FH2} is also a consequence of the equivalence \eqref{ODainv}.
Regarding a scalar function $f(\hat\bsx \cdot \hat\bsz)$ as a function of $\hat\bsz$,
we may use the spin decomposition \eqref{Pif} to write
\begin{align}
f(\hat\bsx \cdot \hat\bsz)
&=\sum_{m=0}^\infty  \int_{\hat\bsw} d\sig_{D-1} \, 
\Pi_m(\hat\bsz, \hat\bsw) f(\hat\bsx \cdot \hat\bsw)
\nn\\
\label{fkaexpansion}
&=\sum_{m=0}^\infty \dim\bbY^{D}_m \int_{\hat\bsw} d\sig_{D-1} \, 
P^{(D)}_{m}(\hat\bsz\cdot\hat\bsw) f(\hat\bsx \cdot \hat\bsw)\,,
\end{align}
where we used the addition theorem \eqref{projector}. Since each summand is an ${\rm O}(D,\hat\bsx)$-invariant spherical function of $\hat\bsz$ of spin $m$, we may apply the equivalence \eqref{ODainv} to find
\begin{align}
\int_{\hat\bsw} d\sig_{D-1} \, 
P^{(D)}_{m}(\hat\bsz\cdot\hat\bsw) f(\hat\bsx \cdot \hat\bsw)
=\lm_m P^{(D)}_{m}(\hat\bsz\cdot\hat\bsx)\,,
\end{align}
where we introduced
\begin{align}
\lm_m&=\int_{\hat\bsw} d\sig_{D-1} \, 
P^{(D)}_{m}(\hat\bsx\cdot\hat\bsw) f(\hat\bsx \cdot \hat\bsw)
= 
\frac{{\rm vol}(S^{D-2})}{{\rm vol}(S^{D-1})} \int_{-1}^1 dt \, (1-t^2)^{\frac{D-3}{2}} P^{(D)}_{m}(t) f( t )\,.
\end{align}
This concludes the proof of the Funk-Hecke formula \eqref{FH1} with \eqref{FH2}.
\section{Two- and three-point functions}
\setcounter{equation}{0}
\label{sec:correlator}

In this section we introduce helicity representation of conformal correlators with symmetric traceless tensors in general spacetime dimension $d$. After elaborating on the helicity decomposition of spinning operators, we utilize the Funk-Hecke formula to derive helicity representation of two- and three-point functions in momentum space\footnote{
See, e.g.,~\cite{Ferrara:1974nf,Sotkov:1976xe,Sotkov:1980qh,Antoniadis:2011ib,Maldacena:2011nz,Coriano:2012hd,Chowdhury:2012km,Bzowski:2012ih,Mata:2012bx,Coriano:2013jba,Bzowski:2013sza,Huh:2013vga,Ghosh:2014kba,Kundu:2014gxa,Arkani-Hamed:2015bza,Kundu:2015xta,Sen:2015doa,Bzowski:2015pba,Bzowski:2015yxv,Jacobs:2015fiv,Myers:2016wsu,Lucas:2016fju,Lucas:2017dqa,Bzowski:2017poo,Coriano:2018bbe,Bzowski:2018fql,Gillioz:2018mto,Coriano:2018bsy,Albayrak:2018tam,Arkani-Hamed:2018kmz,Skvortsov:2018uru,Farrow:2018yni,Isono:2019ihz,Maglio:2019grh,Sleight:2019mgd,Sleight:2019hfp,Albayrak:2019yve} for related developments on conformal correlators in momentum space.}. We then discuss their analytic properties. The cubic vertices introduced there will be used in the next section to construct the crossing symmetric basis.

\subsection{Helicity decomposition of spinning operators}

A standard technique to handle a symmetric traceless tensor in CFT is to contract all vector indices in the tensor operator with a null vector $\bsep$ called the polarization vector \cite{Bargmann:1977gy,Costa:2011mg}\footnote{We use the bold and ordinary fonts for vectors and their components, e.g., $\bsep$ and $\ep^\mu$. Here $\mu$ is the vector index.}. More explicitly, we denote this in a shorthand notation,
\begin{align}
\ep^s.O=\ep^{\mu_1}\ep^{\mu_2}\ldots\ep^{\mu_s}O_{\mu_1\mu_2\ldots\mu_s}\,,
\end{align}
where $s$ is the spin of the operator $O$. In momentum space it is convenient to further decompose the operator by analogy with the helicity decomposition of massless on-shell particles. Without loss of generality, let us use rotational invariance to set
\begin{align}
\bsk=(\bsnl,k)\,,
\end{align}
where $\bsnl$ is the $(d-1)$-component zero vector.
It is then convenient to parameterize the polarization vector as
\begin{align}
\label{ep3pt}
\bsep=(\hat\bsz,i)\,,
\end{align}
where $\hat\bsz$ is a $(d-1)$-component real unit vector.
With this parameterization the contracted tensor operator $\ep^s.O$ can be thought of as a scalar function on the unit sphere $S^{d-2}$ with the coordinate $\hat\bsz$, 
so that we may decompose it into the helicity operators $O_{mn}$ as
\begin{align}
\epsilon^s.O(\bsk)=\sum_{m=0}^s\sum_{n=1}^{\dim\bbY^{d-1}_m}Y_{mn}(\hat\bsz)O_{mn}(\bsk)
\quad{\rm with}
\quad
O_{mn}(\bsk)=\int_{\hat\bsz} d\sigma_{d-2}\,Y_{mn}^*(\hat\bsz)\,\epsilon^s.O(\bsk)\,,
\end{align}
where $\{Y_{mn}\}$ is an orthogonal normal basis for the space of spin $m$ spherical harmonics $\bbY^{d-1}_m$ on $S^{d-2}$. As explained in Sec.~\ref{subsec:FH}, this is the decomposition with respect to the spin $m$ of the little group ${\rm O}(d-1)$ of $\bsk$:
each spin $m$ sector
gives an irreducible representation of the little group.
For later use it is convenient to introduce its conjugate as
\begin{align}
\label{complex_conj-helicity}
\overline{O_{mn}}(-\bsk)=\int_{\hat\bsz} d\sigma_{d-2}\,\bar{\epsilon}^s.O(-\bsk)\,Y_{mn}(\hat\bsz)
\quad
{\rm with}
\quad
\bar{\bsep}=(\hat\bsz,-i)\,,
\end{align}
where note that what is complex conjugate to $O_{mn}(\bsk)$ is not $\overline{O_{mn}}(\bsk)$, but rather $\overline{O_{mn}}(-\bsk)$.
Helicity operators with a general momentum are defined in a similar way by performing an appropriate rotation.

\subsection{Two-point functions}
In momentum space, two-point functions of primary
operators with general spins read \cite{Isono:2018rrb}\footnote{The normalization of the position space correlator is given by 
\begin{align}
\lan \ep_1^s.O(\bsx_1)\bar{\ep}_2^s.O(\bsx_2) \ran 
= \widetilde{C}_{OO} 
\frac{[(\bsep_1\cdot\bar{\bsep}_2)x_{12}^2 - 2(\bsep_1\cdot\bsx_{12})(\bar{\bsep}_2\cdot\bsx_{12})]^s}
{(x_{12}^2)^{\De+s}}
\,,
\end{align}
which is related to our momentum space normalization as
\begin{align}
C_{OO}=2^{s-2\nu}\pi^{\frac{d}{2}}\frac{s!\,\Gamma(-\nu)}{\Gamma(\nu+s+\frac{d}{2})}\widetilde{C}_{OO}\,.
\end{align}}
\begin{align}
\label{V2}
\lan \ep_1^s.O(\bsk)\bar{\ep}_2^s.O(-\bsk) \ran'=
{C}_{OO}  \left(k^2\right)^\nu \! \left[ -\frac{(\bsep_1\!\cdot\!\bsk)(\bar{\bsep}_2\!\cdot\!\bsk)}{k^2} \right]^s
\! P^{(\nu-s, \frac{d}{2}-2)}_s
\bigg( 1-\frac{k^2(\bsep_1\!\cdot\!\bar{\bsep}_2)}{(\bsep_1\!\cdot\!\bsk)(\bar{\bsep}_2\!\cdot\!\bsk)} \bigg)\,, 
\end{align}
where primed correlators are defined by dropping the delta function for momentum conservation as $\lan\,...\,\ran=(2\pi)^d\delta^{(d)}(\sum_i\bsk_i)\lan\,...\,\ran'$. We also introduced $\nu = \De-\frac{d}{2}$ with the scaling dimension $\De$. In this paper we use $\De$ and $\nu+\frac{d}{2}$ interchangeably to simplify equations. $P^{(\alpha, \beta)}_n$ is the Jacobi polynomial defined by
\begin{align}
\label{jacobi-def}
P^{(\al,\bt)}_n(t) = \frac{\Ga(\al+n+1)}{n!\,\Ga(\al+\bt+n+1)} \sum_{m=0}^n 
\binom{n}{m}
\frac{\Ga(\al+\bt+n+m+1)}{\Ga(\al+m+1)} 
\left( \frac{t-1}{2} \right)^m.
\end{align}
Following the last subsection, let us use rotational invariance to set the momentum and the polarization vectors as
\begin{align}
\bsk=(\bsnl,k)\,, \qquad 
\bsep_1=(\hat\bsz_1, i)\,, \qquad \bar{\bsep}_2=(\hat\bsz_2, -i)\,,
\end{align}
where $\hat\bsz_1,\hat\bsz_2$ are $(d-1)$-component real unit vectors. 
The two-point function then becomes
\begin{align}
\label{2ptzeta}
\lan \ep_1^s.O(\bsk)\bar{\ep}_2^s.O(-\bsk) \ran'=
{C}_{OO}  \left(k^2\right)^\nu (-)^s P^{(\nu-s, \frac{d}{2}-2)}_s( -\hat\bsz_1\cdot\hat\bsz_2 )\,,
\end{align}
which is a function of $\hat\bsz_1\cdot\hat\bsz_2$ multiplied by a helicity-independent factor.

\paragraph{Helicity representation.}

We then introduce the helicity representation of the two-point functions.
Since the two-point function \eqref{2ptzeta} is a scalar function of the inner product $\hat\bsz_1\cdot\hat\bsz_2$,
we may apply the Funk-Hecke formula \eqref{FH1} to find
\begin{align}
\lan \ep_1^s.O(\bsk)\bar{\ep}_2^s.O(-\bsk) \ran' 
&=
{C}_{OO}  \left(k^2\right)^\nu \sum_{m=0}^s a_{\nu,s}(m) \,\Pi_{m}(\hat\bsz_1, \hat\bsz_2)
\nn\\
\label{2pt-helbasis}
&=
{C}_{OO} \left(k^2\right)^\nu \sum_{m=0}^s a_{\nu,s}(m)  \sum_{n=1}^{\dim\bbY^{d-1}_m} 
Y_{mn}(\hat\bsz_1) \, Y^*_{mn}(\hat\bsz_2)\,, 
\end{align}
where $\Pi_{m}(\hat\bsz_1, \hat\bsz_2)$ is the projector~\eqref{PiYY} onto the spin $m$ sector $\bbY^{d-1}_m$ on the unit sphere $S_{d-2}$.
The factor $a_{\nu,s}(m)$ is given by the integral formula \eqref{FH2} as
\begin{align}
a_{\nu,s}(m)
&= 
(-)^s \frac{{\rm vol}(S^{d-3})}{{\rm vol}(S^{d-2})} 
\int_{-1}^1 dt \, (1-t^2)^{\frac{d}{2}-2} \, P^{(d-1)}_{m}(t) P^{(\nu-s, \frac{d}{2}-2)}_s( -t )\,,
\end{align}
which is computed in Appendix~\ref{app:am} to find
\begin{align}
\label{am}
a_{\nu,s}(m)
&=
\frac{2^{d-3} \Ga\left(\frac{d-1}{2}\right) \Ga\left(\frac{d}{2}+s-1\right) }{\sqrt{\pi}(s-m)!(s+m+d-3)!}
 \frac{\Ga(d+s-\De-1) \Ga(\De+m-1)}{\Ga(d+m-\De-1) \Ga(\De-1)}\,.
\end{align}
The two-point function of helicity operators now reads
\begin{align}
\lan O_{mn}(\bsk)\,\overline{O_{m'n'}}(-\bsk) \,\ran' 
&=\delta_{mm'}\delta_{nn'}{C}_{OO} \, a_{\nu,s}(m)\left(k^2\right)^\nu\,.
\end{align}

\subsection{Three-point functions of two scalars and one tensor}

We then move on to three-point functions involving two primary scalars and one primary tensor. 
In momentum space they are given by~\cite{Isono:2018rrb}
\begin{align}
\label{V3}
\lan O_1(\bsk_1) O_2(\bsk_2) \ep^s.O(\bsk_3) \ran'
= \sum_{a=0}^s \,(\bsep\cdot\bsk_2)^{s-a} (\bsep\cdot\bsk_3)^a \, V_{12O}^{(a)}(k_1,k_2,k_3)\,,
\end{align}
where we collected a helicity-independent part into the last factor as
\begin{align}
\label{V_123}
&V_{12O}^{(a)}(k_1,k_2,k_3)
\nn\\*
&=
C_{12O}\frac{s!}{a!(s-a)!}\frac{\big(\tfrac{\De+s+\mathfrak{D}_{12O}}{2}-a\big)_{a}}{(\De-1+s-a)_{a}}
\int_0^\infty \!\! 
\frac{dz}{z^{d+1}}z^s\mathcal{B}_{\nu_1}(k_1;z)\mathcal{B}_{\nu_2}(k_2;z)\mathcal{B}_{\nu}(k_3;z)\,.
\end{align}
Here $C_{12O}$ is a normalization factor and $\mathcal{B}_{\nu}$ is defined by
\begin{align}
\label{bulk_boundary_K}
\mathcal{B}_\nu(k;z)&=\frac{1}{2^{\nu-1}\Gamma(\nu)}k^\nu z^{d/2}K_\nu(k_3z)
\end{align}
with $K_\nu(z)$ being the modified Bessel function of the second kind. Note that $\mathcal{B}_{\nu}$ is nothing but the bulk-to-boundary propagator of a {\it scalar} field on $AdS_{d+1}$ with mass $m^2=\nu^2-d^2/4$. 
Also $\mathfrak{D}_{12O}$ is a differential operator defined by ($k_{12}:=|\bsk_1+\bsk_2|$)
\begin{align}
\mathfrak{D}_{12O}=\frac{k_1^2-k_2^2}{k_{12}^2}\left(
k_1\partial_{k_1}+k_2\partial_{k_2}
-\De_t+s+2d
\right)
-\Big[(k_1\partial_{k_1}-\De_1)-(k_2\partial_{k_2}-\De_2)\Big]\,.
\end{align}
We refer the reader to Ref.~\cite{Isono:2018rrb} for details of the helicity independent part~\eqref{V_123}. In the following we instead discuss the helicity structure of the three-point function~\eqref{V3}.

\paragraph{Helicity representation.}

As in the case of two-point functions, we use rotational invariance to fix the momentum of the tensor $O$ and parameterize the polarization vector~as
\begin{alignat}{2}
\bsk_3 &= (\bsnl, k_3) \, , &\qquad \bsep &= (\hat\bsz, i) \, , \label{paramet1}
\end{alignat}
where $\hat\bsz$ is a $(d-1)$-component real unit vector as before. With this parameterization the three-point function~\eqref{V3} reads
\begin{align}
\label{3pt-paramet1}
\lan O_1(\bsk_1) O_2(\bsk_2) \ep^s.O(\bsk_3) \ran'
&= \sum_{a=0}^s 
\left[
 k_2^{s-a} (ik_3)^a \, 
\big( i\cos\te + \hat\bska_2 \cdot \hat\bsz \sin\te \big)^{s-a}
\right]V_{12O}^{(a)}\, ,
\end{align}
where we parameterized the momentum $\bsk_2$ with the $(d-1)$-component unit vector $\hat\bska_2 $ as
\begin{align}
\label{k2}
\bsk_2=k_2(\hat\bska_2\sin\te,\cos\te) \, .
\end{align}
We then decompose the three-point function \eqref{3pt-paramet1} in the spin of the little group of $\bsk_3$.
Since this is a scalar function of the inner product $\hat\bska_2\cdot\hat\bsz$, the Funk-Hecke formula \eqref{FH1} yields the following spin decomposition:
\begin{align}
\label{decomp3pt-paramet1}
\lan O_1(\bsk_1) O_2(\bsk_2) \ep^s.O(\bsk_3) \ran'
&= \sum_{m=0}^s {V}_{12O}(m,\te)\,\Pi^{d-1}_{m}(\hat\bska_2, \hat\bsz)
\nn\\*
&=\sum_{m=0}^s {V}_{12O}(m,\te) \sum_{n=1}^{\dim\bbY^{d-1}_m}
Y^*_{mn}(\hat\bska_2) Y_{mn}(\hat\bsz)
\, .
\end{align}
In other words the three-point function with the helicity operator ${O}_{mn}$ is given by
\begin{align}
\label{helicity_3pt}
\lan O_1(\bsk_1) O_2(\bsk_2) {O}_{mn}(\bsk_3) \ran'
= {V}_{12O}(m,\te)  \, Y^*_{mn}(\hat\bska_2) \, .
\end{align}
Here the $m$-dependent factor ${V}_{12O}(m,\te)$ is given through the integral formula \eqref{FH2} by
\begin{align}
\label{Vmte-paramet1}
{V}_{12O}(m,\te)&= 
\sum_{a=0}^s V^{(a)}_{12O} ~  k_2^{s-a} (ik_3)^a 
\nn\\*
&\quad\quad
\times \frac{{\rm vol}(S^{d-3})}{{\rm vol}(S^{d-2})}
\int_{-1}^1 dt \, (1-t^2)^{\frac{d}{2}-2} P^{(d-1)}_{m}(t) \left[ \, 
\big( i\cos\te + t \sin\te \big)^{s-a}
\right] \, .
\end{align}
In Appendix~\ref{V3m-proof} we derive its analytic expression as
\begin{align}
\label{Vmte-paramet1-analytic}
{V}_{12O}(m,\te)
&= 
\!\sum_{a=0}^{s-m} V^{(a)}_{12O} \,
k_2^{s-a} k_3^a \frac{i^{s-m} (s-a)!}{2^m  (s-a-m)! \left(\frac{d-1}{2}\right)_m}
\sin^m\te \cdot \wh{P}^{(d)}_{s-a,m}(\cos\te) \, ,
\end{align}
where note that the summation is over $0\leq a\leq s-m$.
Here we introduced $\wh{P}^{(d)}_{s-a,m}(t) := {P}^{(d+2m)}_{s-a-m}(t)$. 
For $d=3$, the combination $\sin^m\te \cdot \wh{P}^{(d)}_{s-a,m}(\cos\te)$ is proportional to the associated Legendre function in particular.

\subsection{Analytic properties}

At the end of this section, we discuss analytic properties of three-point functions. In particular, we demonstrate that the non-analytic part of a three-point function enjoys a factorization similarly to scattering amplitudes. Our argument here is parallel to that in Ref.~\cite{Isono:2018rrb}, to which we refer the reader for more detailed explanations.

\medskip
To discuss analytic properties, let us first rearrange Eq.~\eqref{helicity_3pt} into the form,
\begin{align}
\label{helicity_3pt_analyticity}
&\lan O_1(\bsk_1) O_2(\bsk_2) O_{mn}(\bsk_3) \ran'=
(k_2\sin\theta)^m\,Y^*_{mn}(\hat\bska_2) \nn\\
&\qquad\qquad\times
\mathcal{A}_{12O}^{(m)}(\bsk_1,\bsk_2,\bsk_3;\mathfrak{D}_{12O})
\int_0^\infty \!\! 
\frac{dz}{z^{d+1}}z^{s}\mathcal{B}_{\nu_1}(k_1;z)\mathcal{B}_{\nu_2}(k_2;z)\mathcal{B}_{\nu}(k_3;z)\,,
\end{align}
where we introduced a differential operator $\mathcal{A}_{12O}^{(m)}$ as
\begin{align}
\mathcal{A}_{12O}^{(m)}&(\bsk_1,\bsk_2,\bsk_3;\mathfrak{D}_{12O})
=C_{12O}\frac{i^{s-m}\,s!}{2^m  \left(\frac{d-1}{2}\right)_m(s-m)!}
\nn\\*
&\qquad\qquad\quad\times
\sum_{a=0}^{s-m}
 \frac{(s-m)!}{a!(s-m-a)!} \,k_2^{s-m-a} \wh{P}^{(d)}_{s-a,m}(\cos\te) \,
 k_3^{a}\frac{\big(\tfrac{\De+s+\mathfrak{D}_{12O}}{2}-a\big)_{a}}{(\De-1+s-a)_{a}}\,.
\end{align}
Notice that $\mathcal{A}_{12O}^{(m)}$ depends on the helicity only through the spin $m$ of the little group.
An important observation here is that the only source of non-analyticity in the three-point function~\eqref{helicity_3pt_analyticity} is the integral of three bulk-to-boundary propagators $\mathcal{B}_{\nu}(k;z)$: First, the spherical harmonics and the normalized Gegenbauer polynomial are accompanied by an appropriate power in $k_2$ as $(k_2\sin\theta)^m\,Y^*_{mn}(\hat\bska_2)$ and $k_2^{s-m-a}\wh{P}^{(d)}_{s-a,m}(\cos\te)$, so that they are polynomials in $\bsk_2$. Second, the differential operator $\mathfrak{D}_{12O}$ always appears in the form $k_3\mathfrak{D}_{12O}$, which is a polynomial in the momenta $\bsk_i$ and the Euler operators $k_i\partial_{k_i}$ because $\displaystyle(\bsk_2-\bsk_1)\cdot\bsk_3/k_3$ is a component of the vector $\bsk_2-\bsk_1$ along the $\bsk_3$ direction.
Since the Euler operator does not introduce any new non-analyticity, we conclude that the only source of non-analyticity is the integral in Eq.~\eqref{helicity_3pt_analyticity}.

\medskip
As we mentioned, the integrand of Eq.~\eqref{helicity_3pt_analyticity} contains three bulk-to-boundary propagators $\mathcal{B}_{\nu}(k;z)$. By looking at analytic properties of $\mathcal{B}_{\nu}(k;z)$, we find that the integral enjoys the following factorization (see Sec.~3 of Ref.~\cite{Isono:2018rrb} for details):
\begin{align}
&{\rm Disc}_{k_3^2}\int_0^\infty \frac{dz}{z^{d+1}}z^s\mathcal{B}_{\nu_1}(k_1;z)\mathcal{B}_{\nu_2}(k_2;z)\mathcal{B}_{\nu_3}(k_3;z)
\nn\\*
\label{triple-K_n.a.}
&=-\frac{\Gamma(1-\nu_3)}{2^{\nu_3}}\int_0^\infty \frac{dz}{z^{d+1}}z^s\mathcal{B}_{\nu_1}(k_1;z)\mathcal{B}_{\nu_2}(k_2;z)z^{d/2}k_3^{-\nu_3}I_{\nu_3}(k_3z)\times{\rm Disc}_{k_3^2}\left(k_3^2\right)^{\nu_3}\,,
\end{align}
where ${\rm Disc}_{z}$ denotes a discontinuity on the complex $z$ plane. Also, $I_\nu(z)$ is the modified Bessel function of the first kind. Since the prefactor and the differential operator $\cA_{12O}^{(m)}$ in Eq.~\eqref{helicity_3pt_analyticity} do not generate any new non-analyticity as discussed above, the non-analytic part of the three-point function~\eqref{helicity_3pt_analyticity} also factorizes as
\begin{align}
\label{factorization_3pt}
\text{Disc}_{k_3^2}\langle O_1(\bsk_1)O_2(\bsk_2)O_{mn}(\bsk_3)\rangle'
&=T_{12;O_{mn}}(\bsk_1,\bsk_2;\bsk_3)\text{Disc}_{k_3^2}\langle \overline{O_{mn}}(-\bsk_3)O_{mn}(\bsk_3)\rangle'
\,,
\end{align}
where we introduced what we call the cubic vertex $T_{12;O_{mn}}$ as
\begin{align}
&T_{12;O_{mn}}(\bsk_1,\bsk_2;\bsk_3)=
(k_2\sin\theta)^m\,Y^*_{mn}(\hat\bska_2) \nn\\
&\qquad\quad\times
\frac{-\Gamma(1-\nu_3)}{2^{\nu_3}}\frac{\cA_{12O}^{(m)}}{C_{OO}a_{s,\nu_3}(m)}
\int_0^\infty \frac{dz}{z^{d+1}}\mathcal{B}_{\nu_1}(k_1;z)\mathcal{B}_{\nu_2}(k_2;z)z^{d/2}k_3^{-\nu_3}I_{\nu_3}(k_3z)\,.
\label{cubic_V_spin}
\end{align}
Notice that the cubic vertex is analytic at $k_3=0$. We will use it in the next section to construct a crossing symmetric basis for scalar four-point functions.

\paragraph{Three-point functions with a conjugate operator.}
For later use, it is convenient to write down three-point functions involving the conjugate operator $\overline{O_{mn}}$ explicitly. Let us first recall that
\begin{align}
\lan O_1(-\bsk_1) O_2(-\bsk_2) \overline{O_{mn}}(-\bsk_3) \ran'=\Big(\,\lan O_1(\bsk_1) O_2(\bsk_2) O_{mn}(\bsk_3) \ran'\,\Big)^*\,.
\end{align}
Correspondingly, the factorization~\eqref{factorization_3pt} has a conjugate counterpart,
\begin{align}
\nonumber
&\text{Disc}_{k_3^2}\langle O_1(-\bsk_1)O_2(-\bsk_2)\overline{O_{mn}}(-\bsk_3)\rangle' \nn\\
&\qquad
=\Big(\,T_{12;O_{mn}}(\bsk_1,\bsk_2;\bsk_3)\text{Disc}_{k_3^2}\langle \overline{O_{mn}}(-\bsk_3)O_{mn}(\bsk_3)\rangle'\,\Big)^*
\nn\\
\label{factorization_conjugate}
&\qquad=\overline{T_{12;O_{mn}}}(-\bsk_1,-\bsk_2;-\bsk_3)
\text{Disc}_{k_3^2}\langle O_{mn}(\bsk_3)\overline{O_{mn}}(-\bsk_3)\rangle'
\,,
\end{align}
where we introduced the conjugate cubic vertex $\overline{T_{12;O_{mn}}}$ as
\begin{align}
&\overline{T_{12;O_{mn}}}(-\bsk_1,-\bsk_2;-\bsk_3)
\nn\\
&:=\Big(\,T_{12;O_{mn}}(\bsk_1,\bsk_2;\bsk_3)\,\Big)^*
\nn\\
&=
(k_2\sin\theta)^m\,Y_{mn}(\hat\bska_2)
\frac{-\Gamma(1-\nu_3)}{2^{\nu_3}}
\nn\\
\label{T_conjugate}
&\quad\times
\frac{\Big(\,\cA_{12O}^{(m)}(\bsk_1,\bsk_2,\bsk_3;\mathfrak{D}_{12O})\,\Big)^*}{C_{OO}a_{s,\nu_3}(m)}
\int_0^\infty \frac{dz}{z^{d+1}}\mathcal{B}_{\nu_1}(k_1;z)\mathcal{B}_{\nu_2}(k_2;z)z^{d/2}k_3^{-\nu_3}I_{\nu_3}(k_3z)\,.
\end{align}
Note that $\theta$ and $\hat\bska_2$ are defined in Eq.~\eqref{k2} in the frame satisfying Eq.~\eqref{paramet1}.

\section{Crossing symmetric basis for scalar four-point functions}
\setcounter{equation}{0}

\label{sec:crossing}

We then construct the crossing symmetric basis for scalar four-point functions using the ingredients introduced in the previous section. After reviewing the strategy employed in our previous work~\cite{Isono:2018rrb}, we provide an explicit construction in general dimension $d$.

\subsection{Strategy}
\label{subsec:review}

In the previous section we have demonstrated that the non-analytic part of three-point functions factorizes into the cubic vertex~\eqref{cubic_V_spin} and the two-point function as
\begin{align}
\text{Disc}_{k_3^2}\langle O_1(\bsk_1)O_2(\bsk_2)O_{mn}(\bsk_3)\rangle'
&=T_{12;O_{mn}}(\bsk_1,\bsk_2;\bsk_3)\text{Disc}_{k_3^2}\langle \overline{O_{mn}}(-\bsk_3)O_{mn}(\bsk_3)\rangle'\,,
\end{align}
which is analogous to on-shell factorization of scattering amplitudes\footnote{Note that three-point functions enjoy factorization because our correlators are not amputated.}. Its conjugate counterpart is given in Eq.~\eqref{factorization_conjugate} with the conjugate cubic vertex~\eqref{T_conjugate}. Similarly, we require that the non-analytic part of four-point functions factorizes as
\begin{align}
\nonumber
&{\rm Disc}_{\mathbf{s}}\langle O_1(\bsk_1)O_2(\bsk_2)O_3(\bsk_3)O_4(\bsk_4)\rangle'
\\
\label{s_factorization}
&=\sum_{O}\sum_{m,n}
T_{12;O_{mn}}(\bsk_1,\bsk_2;-\bsk_{12}) ~
{\rm Disc}_{\textbf s}\lan \ol{O_{mn}}(\bsk_{12})O_{mn}(-\bsk_{12}) \ran' ~
\ol{T_{34;O_{mn}}}(\bsk_3,\bsk_4;\bsk_{12})\,,
\end{align}
where the first sum is over all the intermediate primary operators and the second is over helicity components $O_{mn}$ of the operator $O$. We also introduced  $\bsk_{ij}=\bsk_i+\bsk_j$ and the Mandelstam type variables,
\begin{align}
\mathbf{s}
=-(\bsk_1+\bsk_2)^2\,,
\quad
\mathbf{t}=-(\bsk_1+\bsk_3)^2\,,
\quad
\mathbf{u}=-(\bsk_1+\bsk_4)^2\,.
\end{align}
Since the on-shell conditions are not imposed on the external momenta, these variables are independent in contrast to the scattering amplitude case. We require similar factorization in the other channels as well.

\medskip
Based on the factorization property, we introduce a crossing symmetric basis for conformal four-point functions as~\cite{Polyakov:1974gs}
\begin{align}
\langle O_1(\bsk_1)O_2(\bsk_2)O_3(\bsk_3)O_4(\bsk_4)\rangle'
=\sum_O\left(W_O^{(\mathbf{s})}+W_O^{(\mathbf{t})}+W_O^{(\mathbf{u})}\right)+\text{(analytic terms)}\,,
\end{align}
where the second term stands for analytic terms which cannot be determined only from analyticity\footnote{
To constrain the analytic contributions, other ingredients will be necessary such as consistency with OPE or locality of the dual bulk theory. Note that the analytic terms correspond to bulk contact terms. }. The label $O$ again runs over all the intermediate primary operators. The function $W_O^{(\mathbf{s})}$, which we call the $\mathbf{s}$-channel Polyakov block, is a conformally covariant function enjoying the following two properties:
\begin{enumerate}
\item $W_O^{(\mathbf{s})}$ reproduces the $\mathbf{s}$-channel factorization:
\begin{align}
\nonumber
{\rm Disc}_{\mathbf{s}}W_O^{(\mathbf{s})}&=\sum_{m,n}
T_{12;O_{mn}}(\bsk_1,\bsk_2;-\bsk_{12})
\\
\label{criterion1}
& \qquad\quad\times
{\rm Disc}_{\textbf s}\lan \ol{O_{mn}}(\bsk_{12})O_{mn}(-\bsk_{12}) \ran' ~
\ol{T_{34;O_{mn}}}(\bsk_3,\bsk_4;\bsk_{12})\,.
\end{align}

\item $W_O^{(\mathbf{s})}$ has no non-analyticity other than the one required by $\mathbf{s}$-channel factorization. In particular, it is analytic in $\bsk_{13}$ and $\bsk_{14}$, so free from $\mathbf{t},\mathbf{u}$-channel discontinuity.
\end{enumerate}
Also, $W_O^{(\mathbf{t})}$ and $W_O^{(\mathbf{u})}$ are $\mathbf{t},\mathbf{u}$-channel analogues of $W_O^{(\mathbf{s})}$ and enjoy similar properties. This basis manifests the crossing symmetry whereas the consistency with the OPE is obscured, hence the latter provides a nontrivial constraint on the theory~\cite{Polyakov:1974gs,Sen:2015doa}.

\subsection{Construction of Polyakov block}
\label{subsec:Polykov}

Let us proceed to constructing the $\textbf s$-channel Polyakov block. In our previous work~\cite{Isono:2018rrb} we have shown that the Polyakov block with an intermediate scalar operator is nothing but the scalar-exchange Witten diagram\footnote{See~\cite{Gopakumar:2016wkt,Gopakumar:2016cpb} for construction in the Mellin space.}:
\begin{align}
W_O^{(\mathbf{s})}&=C_{12O}C_{34O}\int_0^\infty \frac{dz_1}{z_1^{d+1}}\int_0^\infty \frac{dz_2}{z_2^{d+1}}
\nn\\*
&\quad
\times
\mathcal{B}_{\nu_1}(k_1;z_1)\mathcal{B}_{\nu_2}(k_2;z_1)
\cG_{\nu_O}(k_{12};z_1,z_2)
\mathcal{B}_{\nu_3}(k_3;z_2)\mathcal{B}_{\nu_4}(k_4;z_2)\,,
\end{align}
where we introduced the bulk-to-bulk propagator of the would-be dual bulk scalar as
\begin{align}
\cG_\nu(k;z_1,z_2)
&=\frac{\Gamma(1-\nu)\Gamma(1-\nu)}{C_{OO}\,2^{2\nu}}z_1^{d/2}z_2^{d/2}
\Big[I_\nu(kz_1)I_\nu(kz_2)
\nn\\
&\qquad\qquad
-\theta(z_1-z_2)I_{-\nu}(kz_1)I_\nu(kz_2)
-\theta(z_2-z_1)I_{\nu}(kz_1)I_{-\nu}(kz_2)
\Big]
\nn\\
\label{bulk-bulk-scalar}
&=
\frac{-\Gamma(1-\nu)}{C_{OO}\,2^{2\nu-1}\Gamma(\nu)}
\left[
\theta(z_1-z_2)z_1^{d/2}z_2^{d/2}K_\nu(kz_1)I_\nu(kz_2)+(1\leftrightarrow 2)
\right]\,.
\end{align}
The non-analytic part of the bulk-to-bulk propagator, i.e., the first line of Eq.~\eqref{bulk-bulk-scalar}, is responsible for the factorization so that the first criterion~\eqref{criterion1} of the Polyakov block is satisfied. Furthermore, the last expression of Eq.~\eqref{bulk-bulk-scalar} guarantees that the bulk-to-bulk propagator exponentially dumps down for large $z_i$, hence there appear no undesirable singularities, e.g., in the collinear limit $k_1+k_2=k_{12}$. These criteria specify the form of the Polyakov block up to analytic terms, which correspond to bulk contact interactions. See~\cite{Isono:2018rrb} for more details.

\medskip
The key observation for extending the construction for intermediate scalars to general spinning operators is that the differential operators,
\begin{align}
\cA_{12O}^{(m)}(\bsk_1,\bsk_2,-\bsk_{12};\mathfrak{D}_{12O})
\quad\text{and}\quad
\Big(\,\cA_{34O}^{(m)}(-\bsk_3,-\bsk_4,-\bsk_{12};\mathfrak{D}_{34O})\,\Big)^*\,,
\end{align}
appearing in the cubic vertices~\eqref{cubic_V_spin} and~\eqref{T_conjugate}, do not change the non-analytic properties. As a result, we may easily arrive at
\begin{align}
\label{WsOmn}
& W_{O}^{(\mathbf{s})}
= \sum_{m,n}(k_2\sin\theta_2 \cdot k_4\sin\theta_4)^m\,Y^*_{mn}(\hat\bska_2)Y_{mn}(\hat\bska_4) \nn\\
&\quad\times
\frac{\cA_{12O}^{(m)}(\bsk_1,\bsk_2;-\bsk_{12};\mathfrak{D}_{12O})
\Big(\,\cA_{34O}^{(m)}(-\bsk_3,-\bsk_4,-\bsk_{12};\mathfrak{D}_{34O})\,\Big)^*}{a_{s,\nu_O}(m)} \\
\nn
&\quad\times
\int_0^\infty \!\!\frac{dz_1}{z_1^{d+1-s}}\int_0^\infty \!\!\frac{dz_2}{z_2^{d+1-s}}
\mathcal{B}_{\nu_1}(k_1;z_1)\mathcal{B}_{\nu_2}(k_2;z_1)
\mathcal{G}_{\nu_O}(k_{12};z_1,z_2)
\mathcal{B}_{\nu_3}(k_3;z_2)\mathcal{B}_{\nu_4}(k_4;z_2) \,,
\end{align}
where $\mathcal{G}_{\nu_O}(k_{12};z_1,z_2)$ is the {\it scalar} bulk-to-bulk propagator defined by Eq.~\eqref{bulk-bulk-scalar}. We also defined the angles $\theta_{2,4}$ and the unit vectors $\hat\bska_{2,4}$ in the frame $-\bsk_{12}=(\bsnl,k_{12})$ such that\footnote{
Our definition of $\theta_4$ and $\hat\kappa_4$ is motivated by the fact that $\ol{T_{34;O_{mn}}}(\bsk_3,\bsk_4;\bsk_{12})$ is conjugate to $T_{34;O_{mn}}(-\bsk_3,-\bsk_4;-\bsk_{12})$.}
\begin{align}
\bsk_2=k_2(\hat\bska_2\sin\te_2, \cos\te_2) \,, \qquad
-\bsk_4=k_4(\hat\bska_4\sin\te_4, \cos\te_4) \,\,.
\end{align}
Again, the non-analytic part of the propagator is responsible for the $\mathbf{s}$-channel factorization and its large $z_i$ behavior guarantees that there are no undesirable singularities.

\medskip
Furthermore, since in Eq.~\eqref{WsOmn} the spherical harmonics in the first line are the only $n$-dependent factors, the sum over $n$ is nothing but the addition theorem \eqref{projector} with \eqref{PiYY}. The result is  
\begin{align}
& W_{O}^{(\mathbf{s})}
= \sum_{m=0}^s\dim\bbY^{d-1}_m \cdot (k_2\sin\theta_2 \cdot k_4\sin\theta_4)^m\,
P^{(d-1)}_m(\hat\bska_2\cdot\hat\bska_4) \nn\\
&\quad\times
\frac{\cA_{12O}^{(m)}(\bsk_1,\bsk_2;-\bsk_{12};\mathfrak{D}_{12O})
\Big(\,\cA_{34O}^{(m)}(-\bsk_3,-\bsk_4,-\bsk_{12};\mathfrak{D}_{34O})\,\Big)^*}{a_{s,\nu_O}(m)} \\
\nn
&\quad\times
\int_0^\infty \!\!\frac{dz_1}{z_1^{d+1-s}}\int_0^\infty \!\!\frac{dz_2}{z_2^{d+1-s}}
\mathcal{B}_{\nu_1}(k_1;z_1)\mathcal{B}_{\nu_2}(k_2;z_1)
\mathcal{G}_{\nu_O}(k_{12};z_1,z_2)
\mathcal{B}_{\nu_3}(k_3;z_2)\mathcal{B}_{\nu_4}(k_4;z_2) \,.
\end{align}
This concludes our construction of the $\textbf s$-channel Polyakov block with an intermediate operator of an arbitrary spin $s$. The $\textbf{t},\textbf{u}$-channel blocks are defined in a similar fashion.

\section{Conclusion}
\setcounter{equation}{0}
\label{sec:conclusion}

This paper generalized the construction of the crossing symmetric basis of scalar four-point functions in three spacetime dimension \cite{Isono:2018rrb} to general dimensions. To deal with the complication due to spins in general spacetime dimension, we utilized techniques of spherical harmonics.

\medskip
A natural direction to explore along the line of the present work and our previous one~\cite{Isono:2018rrb} is to generalize the construction to correlators involving external spinning operators such as the energy-momentum tensor and other conserved currents. A first step in this direction is the construction of three-point functions with conserved currents and one primary operator of arbitrary spin, which has been done recently in Ref.~\cite{Isono:2019ihz}. There will be no conceptual obstruction to constructing the crossing symmetric basis of four-point functions with external conserved currents based on the results there.

\medskip
Another interesting direction is the extension to de Sitter and inflationary correlators. Some related recent works include \cite{Arkani-Hamed:2018kmz,Sleight:2019mgd,Sleight:2019hfp}. For example, Ref.~\cite{Arkani-Hamed:2018kmz} constructed a crossing symmetric basis of de Sitter four-point functions with external scalars of the conformal mass in four dimensions (dual to scalar operators of conformal weight $\De=2$ in three dimensions) by solving the conformal Ward-Takahashi identities and studying (non-)analytic properties of de Sitter correlators. The extension to four-point functions with external massless scalars was also explored there aiming at applications to inflationary physics. More recently, Refs.~\cite{Sleight:2019mgd,Sleight:2019hfp} developed a Mellin representation of exchange diagrams on (anti-)de Sitter space in momentum space for external scalars with arbitrary mass. It would be interesting to explore the relation of these works with ours. We hope to revisit these issues in the near future.

\section*{Acknowledgments}
We would like to thank Toshiaki Takeuchi for useful discussion and related collaborations. HI is supported by the ``CUniverse'' research promotion project by Chulalongkorn University (grant reference CUAASC). TN is supported in part by JSPS KAKENHI Grant Numbers JP17H02894 and JP18K13539, and MEXT KAKENHI Grant Number JP18H04352. GS is supported in part by the DOE grant DE-SC0017647 and the Kellett Award of the University of Wisconsin.

\appendix

\section{Derivation of Eq.~(\ref{am})}
\label{app:am}
We derive the analytic expression \eqref{am} of the factor $a_{\nu,s}(m)$. We present the integral form of $a_{\nu,s}(m)$ again,
\begin{align}
a_{\nu,s}(m)
&= 
(-)^s \frac{{\rm vol}(S^{d-3})}{{\rm vol}(S^{d-2})} 
\int_{-1}^1 dt \, (1-t^2)^{\frac{d}{2}-2} P^{(d-1)}_{m}(t) P^{(\nu-s, \frac{d}{2}-2)}_s( -t )\,.
\end{align}
To evaluate this, it is convenient to notice that the Gegenbauer polynomial is a special case of the Jacobi polynomial:
\begin{align}
C_m^{(\lambda)}(t)=(-)^mC_m^{(\lambda)}(-t)=\frac{(2\lambda)_m}{(\lambda+\frac{1}{2})_m}P_m^{(\lambda-\frac{1}{2},\lambda-\frac{1}{2})}(t)\,,
\end{align}
which may be rephrased in terms of the normalized Gegenbauer polynomial~\eqref{normalized_G} as
\begin{align}
P^{(D)}_{m}(t)=(-)^mP^{(D)}_{m}(-t)=\frac{m!}{(\frac{D-1}{2})_m}P_m^{(\frac{D-3}{2},\frac{D-3}{2})}(t)\,.
\end{align}
By using the integral formula (7.391.10, p.807, \cite{GR})\footnote{This formula is found in the seventh edition of \cite{GR}. A caveat is that the formula in some older editions has a typo. },
\begin{align}
&
\int_{-1}^1 \! dt \, (1-t)^\al (1+t)^\bt  P^{(\al,\bt)}_m(t) P^{(\ga,\bt)}_n(t) \nn\\
& \quad
= 
\frac{2^{\al+\bt+1} ~ \Ga(\ga+\bt+n+1+m) \Ga(\ga-\al+n-m) \Ga(\al+m+1) \Ga(\bt+n+1)}
{m!(n-m)! ~ \Ga(\ga+\bt+n+1) \Ga(\ga-\al) \Ga(\al+\bt+m+n+2)}\,,
\label{jacobi-int}
\end{align}
we obtain
\begin{align}
\label{lm2pt}
a_{\nu,s}(m)
&=
\frac{2^{d-3} \Ga\left(\frac{d-1}{2}\right) \Ga\left(\frac{d}{2}+s-1\right) }{\sqrt{\pi}(s-m)!(s+m+d-3)!}
 \frac{\Ga(d+s-\De-1) \Ga(\De+m-1)}{\Ga(d+m-\De-1) \Ga(\De-1)}\,.
\end{align}

\section{Derivation of Eq.~(\ref{Vmte-paramet1-analytic})}
\label{V3m-proof}

In this appendix we derive Eq.~\eqref{Vmte-paramet1-analytic} by performing the integral \eqref{Vmte-paramet1}. We first replace the normalized Gegenbauer polynomial in the integral \eqref{Vmte-paramet1} by its Rodrigues formula
\begin{align}
P^{(d-1)}_{m}(t) = \frac{(-1)^m}{2^m(\frac{d}{2}-1)_m} (1-t^2)^{2-\frac{d}{2}}
\frac{d^m}{dt^m} (1-t^2)^{m+\frac{d}{2}-2} \,,
\end{align}
and integrate the resulting integral by part $m$ times.
Then the integral in the second line of Eq.~\eqref{Vmte-paramet1} becomes
\begin{align}
\frac{{\rm vol}(S^{d-3})}{{\rm vol}(S^{d-2})} 
\frac{i^{s-a-m}(s-a)!\sin^m\te}{2^m(\frac{d}{2}-1)_m(s-a-m)!}
\int_{-1}^1 dt \, (1-t^2)^{m+\frac{d}{2}-2} (\cos\te-it\sin\te)^{s-a-m} \, .
\end{align}
Note that it vanishes when $s-a-m<0$.  
Applying to this integral formula,
\begin{align}
\label{PDm-integral}
P^{(D)}_{m}(\cos\te) = \frac{{\rm vol}(S^{D-3})}{{\rm vol}(S^{D-2})}
\int_{-1}^1 dt \, (1-t^2)^{\frac{D-4}{2}} (\cos\te - it\sin\te)^m \, ,
\end{align}
of the normalized Gegenbauer polynomial derived shortly,
we find the result \eqref{Vmte-paramet1-analytic}.

\medskip

\paragraph{Integral formula of the normalized Gegenbauer polynomial.}
Here we give a derivation of the integral formula \eqref{PDm-integral}, which had already been used in \cite{Isono:2018rrb}. For this, we start from the following integral
\begin{align}
\label{pz}
p(\bsz) := \int_{\hat\bsx} d\sig_{D-2} \, (z_D - i\hat\bsx \cdot \bsz_{(D-1)})^m\,,
\end{align}
where $\bsz$ is a $D$-component vector given by $\bsz=(\bsz_{(D-1)}, z_D)$ with a $(D-1)$-component vector $\bsz_{(D-1)}$ and $\hat\bsx$ is a $(D-1)$-component unit vector.
Since the integrand $(z_D - i\hat\bsx \cdot \bsz_{(D-1)})^m$ is harmonic in $\bsz$, its restriction to $S^{D-1}$, namely $p(\hat\bsz)$, is a spherical harmonic of spin $m$\footnote{
Here we used the fact that any spherical harmonic on the unit sphere $S_{D-1}$ is the restriction of a harmonic polynomial in $\mathbb{R}^D$ (i.e. a homogeneous polynomial annihilated by the Laplacian in $\mathbb{R}^D$) to $S_{D-1}$. The spin of the spherical harmonic is the degree of homogeneity of the corresponding harmonic polynomial.
}.
Furthermore, this is ${\rm O}(D,\hat\bse_D)$-invariant, where $\hat\bse_D=(\bsnl,1)$ with $(D-1)$-component zero vector $\bsnl$.
We may therefore apply the theorem \eqref{ODainv} to the spherical harmonic $p(\hat\bsz)$ of spin $m$, to find
\begin{align}
\label{PDm}
p(\hat\bsz) = P^{(D)}_{m}(\hat\bsz \cdot \hat\bse_D) \,,
\end{align}
where the overall normalization factor is 1 because $p(\hat\bse_D)=1$.
Combining Eq.~\eqref{PDm} with Eq.~\eqref{pz}, we find the desired integral representation of the normalized Gegenbauer polynomial,
\begin{align}
P^{(D)}_{m}(\cos\te) = \frac{{\rm vol}(S^{D-3})}{{\rm vol}(S^{D-2})}
\int_{-1}^1 dt \, (1-t^2)^{\frac{D-4}{2}} (\cos\te - it\sin\te)^m \,,
\end{align}
where we introduced the angle $\te$ by $z_D=|\bsz|\cos\te$.

\bibliography{crossing-genD}{}
\bibliographystyle{utphys}

\end{document}